\documentclass[aip,reprint,letterpaper,noeprint,longbibliography,nodoi,footinbib]{revtex4-1}

\usepackage[colorlinks, allcolors=blue]{hyperref}
\usepackage{amsmath}
\usepackage{amsfonts}
\usepackage{graphicx}
\usepackage{float}
\usepackage{color}

\newcommand{\beq} {\begin{equation}}
\newcommand{\eeq} {\end{equation}}
\newcommand{\bal} {\begin{aligned}}
\newcommand{\eal} {\end{aligned}}

\begin{document}

\title{
From Rock Tumbling to Telescope Mirror Grinding: \\
How Optical Perfection Emerges from Random Motion
}

\author{Emmanuel Schaan}
\affiliation{SLAC National Accelerator Laboratory, 2575 Sand Hill Road, Menlo Park, California 94025, USA}
\affiliation{Kavli Institute for Particle Astrophysics and Cosmology, Stanford University, 452 Lomita Mall, Stanford, CA, 94305, USA}

\begin{abstract}

When two glass disks are rubbed together by hand with an abrasive between them, their surfaces spontaneously converge to perfectly matching spherical curves, accurate to tens of nanometers.
How does such optical precision emerge from an essentially random manual process?
Starting from Preston's law of material removal, 
we derive a first-principles explanation.
Symmetries of the grinding stroke imply that planes and spheres are the generic equilibria of the process.
Their stability is governed by a hierarchy of timescales: small-scale irregularities decay exponentially on the grinding timescale $T_\text{grind}$, while large-scale deviations from a sphere diffuse away on the longer spherizing timescale $T_\text{sphere} \propto (R_\text{mirror}/R_\text{stroke})^2\, T_\text{grind}$.
Gravity and finite disk size destabilize planar surfaces, but for sufficiently short strokes spheres remain stable
($T_\text{deform} \gg T_\text{sphere}$), with only their radius of curvature varying adiabatically.
The same framework unifies telescope mirror grinding with everyday phenomena: rock tumbling (explaining why rotary tumbling rounds edges more than vibratory, and why neither produce spheres) and sanding (explaining why small flat surfaces inevitably round).
These examples offer an accessible classroom context for stable and unstable equilibria, exponential relaxation, diffusion, multi-timescale analysis, and the emergence of the Laplacian operator from averaging over a local area.

\end{abstract}

\maketitle

\section{Introduction}

\begin{figure*}[!htbp]
\centering
\includegraphics[width=\textwidth]{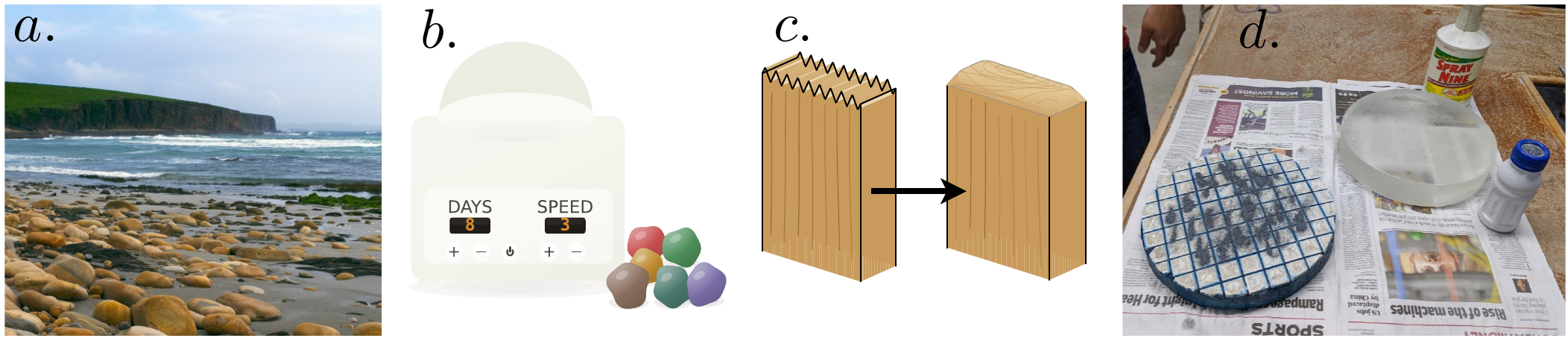}
\caption{
The same physics responsible for (a)~polished cobbles on a beach (credit: Pix Tresa, Free to use under the Unsplash License), (b)~tumbled rocks, and (c)~for the unwanted rounded edges when sanding small objects can be used to (d)~make perfectly smooth and spherical optical surfaces like telescope mirrors (credit: the author), with no more than one's bare hands.
Sec.~\ref{sec:rocks} explains why tumbling and sanding yield smooth but non-spherical surfaces, while mirror grinding converges to a sphere.
}
\label{fig:analogy_mirror_rock_tumbling_sanding}
\end{figure*}

Polished river cobbles, tumbled rocks, and the frustrating tendency of sanding to round small surfaces all share a common physics with one of the most elegant feats of amateur craftsmanship: grinding a telescope mirror by hand to optical precision.
In all three cases, surfaces are rubbed together with an abrasive, and the differing outcomes are determined by differing hierarchies of timescales (Fig.~\ref{fig:analogy_mirror_rock_tumbling_sanding}).

A glass blank can be ground and polished by hand into a perfect spherical mirror for a telescope.
The process involves rubbing two matching glass disks, the future mirror and the tool, in somewhat random directions with progressively finer abrasive grit.
Extremely simple strokes are performed by hand (Fig.~\ref{fig:grinding_strokes}).
Yet remarkably, the two surfaces spontaneously converge to perfectly matching, perfectly spherical curves, accurate to a few tens of nanometers.
How is this possible?

Excellent references present the amateur telescope making process in detail: the classic manuals~\cite{texereau1984, rutten1988, ingalls1953, Cook1995, Berry2001, Lecleire2006, Thompson1985, Howard1984}, the online and video resources popular among practitioners~\cite{StellafaneTM, BBastro, LOptics, Jacquin, WaiteYouTube, Astrokhan},
and the professional fabrication literature focus on practical methods and reproducibility \cite{anderson2001, rupp1965, rupp1972, suratwala2014convergent}.
Yet neither provides a first-principles physical explanation for \emph{why} the process works.
Ref.~\cite{rupp1965} addresses the ``maintenance problem'' (how spherical surfaces remain spherical) rather than the ``convergence problem'' (why arbitrary surfaces become spherical in the first place).
Refs.~\cite{rupp1972, suratwala2014convergent} identify empirical conditions or state convergence qualitatively, without derivation.
The rock tumbling literature \cite{krumbein1941, firey1974, bloore1977, domokos2012, havasi2023, priour2020} offers useful clues: ref.~\cite{krumbein1941} identifies a separation of timescales empirically, with smoothing preceding spherizing.
The present paper is thus novel in three ways:
it derives, rather than assumes, the convergence of hand grinding towards matching spheres; it does so for the two-body mirror and tool problem, whose matching dynamics are not present in the single-body pebble literature; and it unifies mirror grinding, rock tumbling and sanding as regimes of the same equations, distinguished only by their hierarchies of timescales.

In this paper, we provide the missing first-principles explanation, starting from Preston's law of abrasion \cite{preston1927}.
We show that symmetries of the grinding motion make planes and spheres generic equilibria of the process, and that a hierarchy of timescales governs how surfaces smooth, match, and spherize.
The same framework unifies mirror grinding, rock tumbling, and sanding, explaining their differing outcomes.

This study offers an original context to study stable and unstable equilibria, exponential relaxation, diffusion, multi-timescale analysis, and the emergence of the Laplacian from spatial averaging.
This could be done in an upper-division optics, mechanics or mathematical methods class.
Observing river rocks, rock tumbling and wood sanding are all activities that can be performed in a lab class.
Grinding a small (e.g., 6'') telescope mirror by hand can be performed by a student in about 20 hours as part of an astronomy club (e.g., Chabot Space and Science Center in Oakland, CA{\footnote{Chabot Space and Science Center, Oakland, CA, \url{https://chabotspace.org/}.}}).

\section{Amateur telescope mirror grinding: the empirical recipe}

\begin{figure}[!htbp]
\centering
\includegraphics[width=0.8\columnwidth]{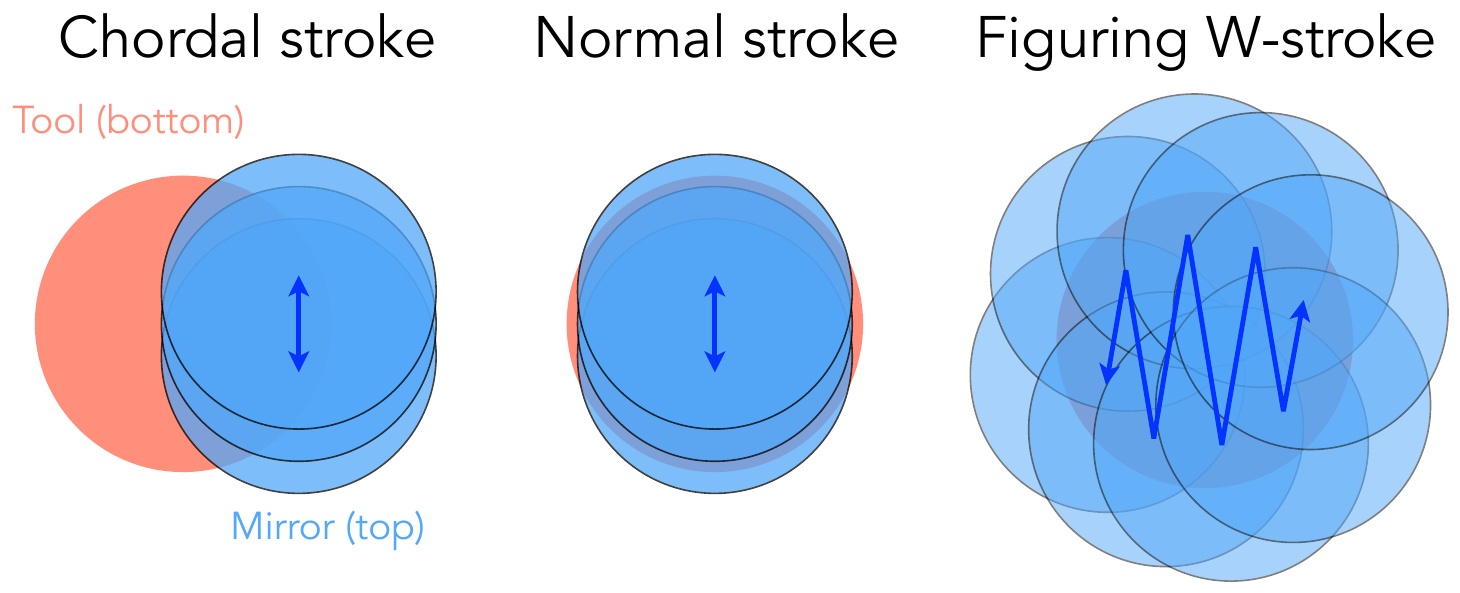}
\caption{
Three typical grinding strokes.
The chordal stroke (left), with a large overhang, allows to quickly dig the mirror center and the tool's edges, but leads to aspheric, mismatching surfaces.
The ``normal stroke (center) with $\simeq 1/3$~diameter full travel, allows to spherize and match the two surfaces.
The W-stroke (right) allows to smoothly deviate from a sphere towards a paraboloid in the phase called figuring.
In all cases, the tool and mirror are slowly rotated to sample all possible directions.
This paper explains how such simple, approximate and somewhat random rubbing motions can lead to perfect optical surfaces.
}
\label{fig:grinding_strokes}
\end{figure}

In practice, the operator typically stands at a table where the tool disk has been fixed and 
the mirror disk placed on top, pressing down with both hands.
Each stroke is a short back-and-forth translation, roughly along the operator's body axis, with a variable length and lateral offset depending on the stroke being performed (Fig.~\ref{fig:grinding_strokes}).
The operator slowly walks around the table and rotates the mirror with respect to the tool, so that all directions are sampled.
The operator first quickly digs the center of the mirror with a chordal stroke, then slowly spherizes it with the normal stroke.
To obtain the desired radius of curvature, the operator alternates mirror on top (MOT), which increases the curvature of the sphere, and tool on top (TOT), which reduces it.
Progressively finer grit is used as the mirror surface becomes smoother, concluding with a final polishing step.
The stroke motion is imprecise and somewhat random, being performed by hand, yet the outcome is perfectly spherical.
No two strokes are perfectly identical, and their directions are sampled roughly uniformly as the operator rotates the disk and walks around the table. 
Each stroke is thus random, but the ensemble of strokes is statistically isotropic; it is this statistical symmetry that underlies the equilibrium surfaces derived below.
The grinding process is long but feasible (Table~\ref{tab:grit_log}).
For Newtonian telescopes, the mirror needs to be made parabolic instead of spherical, a final step called figuring.
This is achieved with the W-stroke, which introduces a minute and smooth deviation from sphericity.

\begin{table}[htbp]
\caption{Time spent at each grit while grinding the author's 10'' f/5 mirror ($\sim 50$~h total).
The first two steps are the longest, taking about 10~hours each: they dig the center (few~mm depth of curve or ``sagitta'') and roughly spherize it.
The subsequent fine-grit stages each take a comparable few hours, rather than scaling with grit size.
Sec.~\ref{sec:numerics} explains all these facts.}
\label{tab:grit_log}
\begin{tabular}{|llc|}
\hline
Stage & Grit size & Duration\\
\hline
Rough shaping: dig the center & 200~$\mu$m (\#80) & 11h50\\
\hline
Rough shaping: spherize & 125~$\mu$m (\#120) & 12h50\\
\hline
Fine grinding: & 68~$\mu$m (\#220) & 5h25\\
smooth the surface & 36~$\mu$m (\#320) & 3h20\\
 & 24~$\mu$m (\#400) & 2h15\\
 & 20~$\mu$m (\#500) & 2h35\\
 & 16~$\mu$m (\#600) & 2h45\\
 & 12~$\mu$m & 2h40\\
 & 9~$\mu$m & 2h20\\
 & 5~$\mu$m & 2h25\\
\hline
Total & & $\sim 50$h\\
\hline
\end{tabular}
\end{table}

Modern professional fabrication automates the same physics through computer-controlled optical surfacing, in which a tool smaller than the mirror, also obeying Preston's law, is driven with a numerically optimized dwell-time map computed from a measured surface-error map~\cite{suratwala2010deterministic, anderson2001, jones1977, aspden1972}. 
That approach is completely deterministic but depends on metrology and feedback, whereas the hand process analyzed here converges to a sphere automatically.

\section{A minimal model of mirror grinding}

\begin{figure}[h]
\centering
\includegraphics[width=0.7\columnwidth]{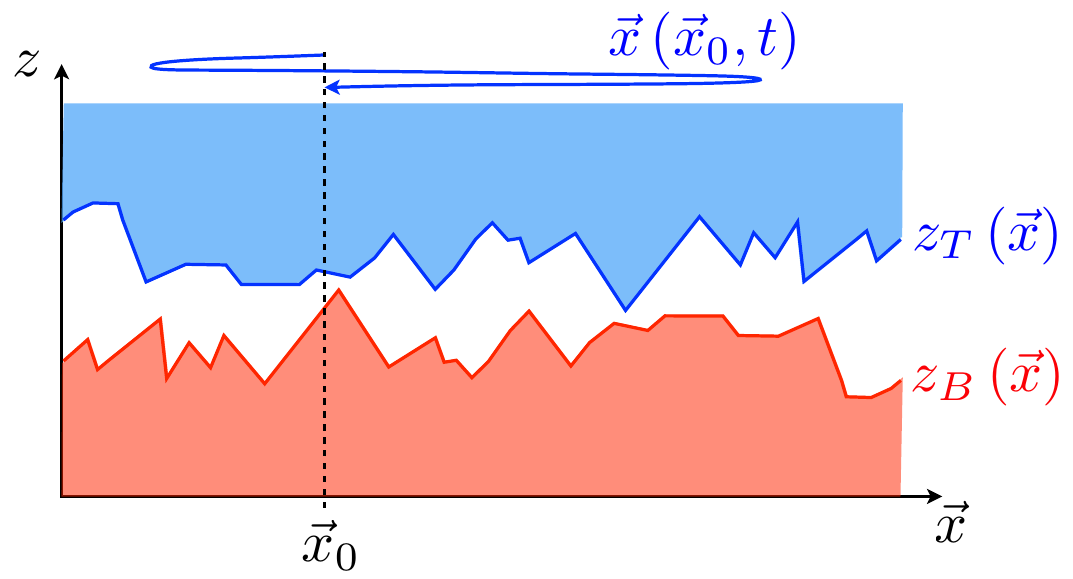}
\caption{
Definition of the top and bottom surface heights ($z_T$ and $z_B$, respectively) and the grinding stroke motion 
$\vec{x} \left( \vec{x}_0, t \right)$.
}
\label{fig:geometry}
\end{figure}

We consider two surfaces pressed against each other, parametrized by their heights $z_T(\vec{x})$ and $z_B(\vec{x})$ over a reference plane as a function of 2D position $\vec{x}$ (Fig.~\ref{fig:geometry}).
During the grinding stroke, the surfaces are pressed together and translated or rotated relative to each other.
The motions allowed are rigid-body motions, i.e. translations and rotations only, since the disks do not deform during the stroke.
We work in the frame of the bottom surface, so that at time $t$, point $\vec{x}_0$ of the bottom surface faces point $\vec{x}(t)$ of the top surface.
The stroke is periodic with period $T_\text{stroke}$, 
i.e. 
$\vec{x}(0) = \vec{x}(T_\text{stroke}) = \vec{x}_0$.

Preston's law \cite{preston1927} states that the material removal rate at each point is proportional to the local pressure $P$ and relative surface speed $v$:
\beq
\partial_t z_B\left( \vec{x}_0, t \right) 
= - \partial_t z_T\left( \vec{x}\left( t \right), t \right)
= -\alpha\ P\left( \vec{x}_0, t \right)
v\left( \vec{x}_0, t \right),
\label{eq:preston}
\eeq
where $\alpha$ (Preston's positive constant, with dimension of an inverse pressure or $m^2/N$) depends on the disk material and abrasive type; typical values are estimated in Sec.~\ref{sec:numerics}.
Physically, harder pressing and faster rubbing both remove material faster; zero pressure or zero relative motion removes none.

For simplicity, we focus on strokes where the relative velocity is roughly spatially uniform{\footnote{For a pure translation the relative velocity is the same at every point of the overlap, so $v$ enters Preston's law as a common prefactor and drives no differential abrasion. For a spinning motion, by contrast, $v$ grows with radius and would grind the edge faster than the center.}} and constant in magnitude (instantaneous direction reversals aside).
This is a good approximation for the strokes in Fig.~\ref{fig:grinding_strokes}.
Under this approximation, the key quantity to model is the pressure field $P(\vec{x})$, which is shaped by three factors: surface irregularities, the finite disk geometry, and the asymmetry between the top and bottom surfaces caused by gravity.
We consider each in turn.

When $z_B(\vec{x}_0)$ and $z_T(\vec{x}(t))$ are in contact, 
one intuitively expects the local pressure to increase with their height difference $z_B(\vec{x}_0) - z_T(\vec{x}(t))$.
Protruding regions (high $z_B$ or low $z_T$) are pressed harder and thus ground faster.
Following this intuition, the elastic (Winkler) foundation model of contact mechanics~\cite{johnson1985, chekina1998} further postulates linearity:
\beq
P = k\,\left( z_B - z_T \right),
\eeq
with stiffness $k$ (units N/m$^3$).
In practice, this simple model is imperfect.
For instance, the pressure field can be non-local: if two surfaces have rare, high peaks, only those peaks touch, while the rest of the surface is unloaded.
We neglect these non-local effects in what follows.

With this simple pressure model, Preston's law \eqref{eq:preston} becomes:
\beq
\boxed{
\bal
&\partial_t z_B\left( \vec{x}_0, t \right) 
= - \partial_t z_T\left( \vec{x}\left( t \right), t \right) \\
&=
- \frac{1}{T_\text{grind}} \left[ z_B\left( \vec{x}_0, t \right) - z_T\left( \vec{x}\left(t\right), t\right) \right],
\label{eq:grinding_toy_model}
\eal
}
\eeq
where $T_\text{grind} \equiv 1/\left( \alpha k v \right)$ absorbs Preston's constant, the pressure stiffness coefficient and the relative surface velocity.
This is the only timescale parameter one can derive from these quantities, and is thus the characteristic timescale of abrasion, estimated in Sec.~\ref{sec:numerics} to be tens of minutes to an hour in practice.
We will find in Sec.~\ref{sec:smooth} that $T_\text{grind}$ is also the e-folding time on which small-scale irregularities decay.
The minus sign can be understood intuitively: if $z_B$ (resp. $z_T$) is high (resp. low), a lot of grinding will occur and $z_B$ (resp $z_T$) will decrease (resp. increase).
All the results in this paper will be derived from Eq.~\eqref{eq:grinding_toy_model}.
Throughout, all timescales denote e-folding times; reducing an amplitude by a factor $F$ takes $\ln F$ of them.

\section{Equilibrium from invariance under the stroke}

An equilibrium configuration is one where the surfaces no longer evolve in time, i.e. $\partial_t z_B = \partial_t z_T = 0$ everywhere.
From Eq.~\eqref{eq:grinding_toy_model}, this requires $z_B(\vec{x}_0) = z_T(\vec{x}(t))$ for all $\vec{x}_0$ and all $t$. 
In particular, this must hold at $t=0$, so the two surfaces must be identical: $z_B = z_T$.

\begin{figure}[htbp]
\centering
\includegraphics[width=0.7\columnwidth]{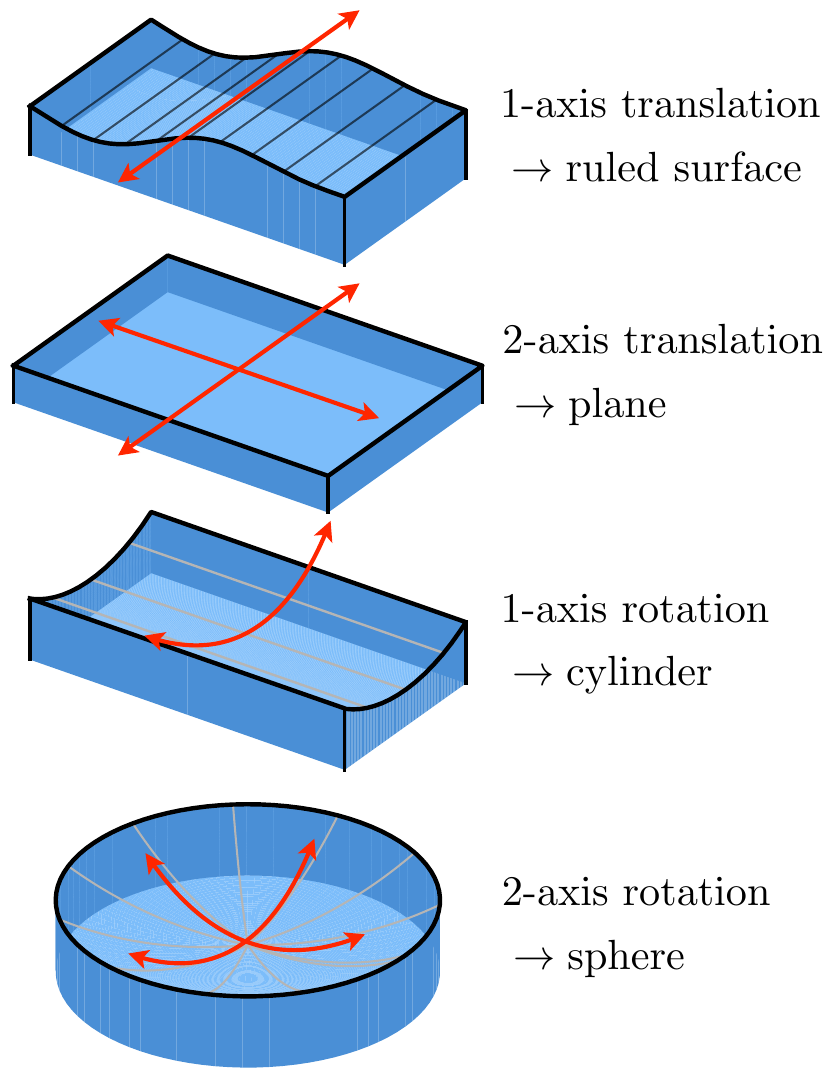}
\caption{
Equilibrium surfaces are those left invariant by the stroke symmetry. From top to bottom: a single translation leaves a surface invariant along that direction (an extruded profile); two independent translations force a plane; a rotation about one axis gives a surface of revolution; rotations about two independent axes force a sphere.
}
\label{fig:symmetries}
\end{figure}

But not every matching pair is an equilibrium.
The condition must hold at all times $t$, which means $z_T(\vec{x}(t)) = z_T(\vec{x}_0)$: the surface must be invariant under all the translations and rotations of the stroke.
Equilibrium configurations are thus precisely those surfaces that share the symmetries of the stroke motion \cite{texereau1984, ingalls1953} (Fig.~\ref{fig:symmetries}):
\begin{itemize}
\item Translation in one direction $\Rightarrow$ surfaces invariant in that direction.
\item Translations in two independent directions $\Rightarrow$ planes.
\item Rotation about a fixed axis $\Rightarrow$ surfaces of revolution (e.g., cylinders, spheres).
\item Rotations about two independent axes $\Rightarrow$ spheres.
\end{itemize}
For generic strokes, then, planes and spheres are the equilibrium configurations.
Interestingly, this result does not strictly require the linear pressure law assumed, and is thus very generic and robust. 
Indeed, any law where the local pressure vanishes if the two surfaces match at this point would result in the same equilibrium configuration.
Whether these equilibria are stable, and thus likely endpoints of the grinding process, requires examining the out-of-equilibrium dynamics.

\section{Stabilizing dynamics: decay and diffusion of surface irregularities}

We focus on the \emph{grinding} regime, $T_\text{stroke} \ll T_\text{grind}$, where the effect of each stroke is small and the surfaces evolve slowly over many stroke periods.
This is the physically relevant limit: a typical stroke takes seconds, whereas a noticeable change in the mirror surface requires minutes to hours.
This is opposite to the \emph{``engraving''} regime, where a single motion dramatically alters the surfaces, like writing one's name on a glass window with a key.

In this regime, we can neglect the explicit time-dependence of $z_T$ and $z_B$ over the short duration of a single stroke.
Indeed, because $T_\text{stroke}\ll T_\text{grind}$, the surfaces barely change during one stroke, so we treat them as frozen over a period. 
Integrating Eq.~\eqref{eq:grinding_toy_model} over one stroke period with frozen surfaces gives the small net change per stroke:
\beq
\bal
\frac{\Delta z_B\left( \vec{x}_0 \right)}
{T_\text{stroke}}
&= 
- \frac{1}{T_\text{grind}} 
\left[ 
z_B\left( \vec{x}_0\right)
- 
\underbrace{\langle z_T\left( \vec{x}\left(t\right)\right) \rangle_\text{stroke}}
_{\text{Spatial average}}
\right]. \\
\eal
\eeq
Summing these increments over many strokes defines a smooth evolution on a coarse-grained time that resolves many strokes but not the motion within one. 
This two-time averaging is the leading order of multiple-scale analysis~\cite{benderorszag1999}.
On the coarse-grained time, the small differences thus become derivatives:
\beq
\boxed{
\partial_t z_B
=
- \frac{1}{T_\text{grind}} 
\left[ 
z_B
- 
\langle z_T\rangle_\text{stroke}
\right]
}
\label{eq:coarse_grained_grinding_dynamics}
\eeq
and by symmetry, the same holds with $z_B \leftrightarrow z_T$.
Thus the stroke motion causes each point of the bottom surface to be ground by the \emph{spatial average} of the top surface over the stroke excursion: a single point interacts with many.

We denote by $R_\text{stroke}$ the typical stroke excursion; for a center-over-center stroke in all directions, this samples a disk of radius $R_\text{stroke}$. 
For the normal $1/3$-diameter stroke, $R_\text{stroke} = R_\text{mirror}/3$.
The effect of the stroke on surface irregularities will depend on the ratio of their spatial scale or wavelength $\lambda$ to $R_\text{stroke}$, as we now examine.

\subsection{Smoothing: decay of small-scale irregularities ($\lambda \ll R_\text{stroke}$)}
\label{sec:smooth}

Small-scale irregularities cancel out in the stroke spatial average $\langle z_T \rangle_\text{stroke}$ of Eq.~\eqref{eq:coarse_grained_grinding_dynamics}. 
This can be understood by considering a moving average (a top-hat filter) of width $R_\text{stroke}$ acting on a short-wavelength mode $z_T(x) = \cos(2\pi x/\lambda)$.
The average is
$\langle \cos(2\pi x/\lambda) \rangle_\text{stroke}
=
\tfrac{1}{R_\text{stroke}}\int_0^{R_\text{stroke}}\cos(2\pi x/\lambda)\,dx$,
and is bounded by 
$|\langle \cos(2\pi x/\lambda) \rangle_\text{stroke}|
\leq
( \lambda/R_\text{stroke}) / (2\pi)
\ll 1
$.

As a result, Eq.~\eqref{eq:coarse_grained_grinding_dynamics} simplifies:
\beq
\partial_t z_B = - \frac{1}{T_\text{grind}} z_B,
\eeq
and identically for $z_T$.
Both surfaces thus decay exponentially, $z_B, z_T \propto e^{-t/T_\text{grind}}$, on the smoothing timescale $T_\text{smooth} \equiv T_\text{grind}$: grinding erases irregularities on scales below the stroke excursion.

\subsection{Matching and spherizing: evolution of large-scale irregularities ($\lambda \gg R_\text{stroke}$)}

\begin{figure}[htbp]
\centering
\includegraphics[width=0.7\columnwidth]{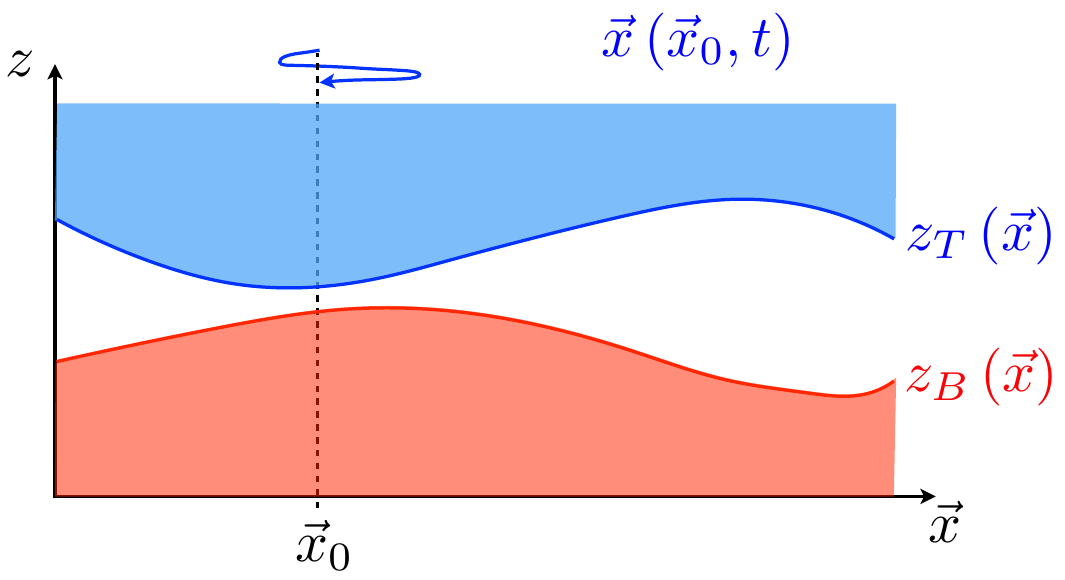}
\caption{
Geometry of surface irregularities varying on scales larger than the stroke radius.
}
\label{fig:geometry_large_scale_diffusion}
\end{figure}

For large-scale irregularities, the stroke samples only a fraction of each peak or trough (Fig.~\ref{fig:geometry_large_scale_diffusion}).
For a stroke radius $R_\text{stroke}$ much smaller than the typical scale $\lambda$ of surface irregularities considered,
the surfaces are well approximated by their Taylor expansion:
\beq
\bal
z_T\!\left( \vec{x} \right)
&= z_T(\vec{x}_0)
+ \left[ \vec{x} - \vec{x}_0 \right] \cdot \vec{\nabla} z_T(\vec{x}_0)\\
&+ \frac{1}{2} \left[ \vec{x} - \vec{x}_0 \right] \cdot \vec{\nabla}\vec{\nabla} z_T(\vec{x}_0) \cdot \left[ \vec{x} - \vec{x}_0 \right]
+ \ldots
\eal
\eeq
For a stroke that uniformly samples a disk of radius $R_\text{stroke}$ centered at $\vec{x}_0$,
the spatial average of this Taylor expansion leads to a Laplacian \cite{styer2015}:
\beq
\bal
\langle z_T\left( \vec{x}\left(t\right)\right) \rangle_\text{stroke}
&=
z_T\left( \vec{x}_0 \right)
+
\frac{R_\text{stroke}^2}{8}\
\nabla^2 z_T\left( \vec{x}_0 \right)\\
&+
\mathcal{O}\left( \left(R_\text{stroke}/\lambda\right)^4 \right).
\eal
\eeq
Indeed, the gradient, cross-terms and higher order odd terms vanish by symmetry, and $\langle (x-x_0)^2\rangle = \langle (y-y_0)^2\rangle = R_\text{stroke}^2/4$.
Intuitively, the Laplacian correction encodes the difference between the surface height at $\vec{x}_0$ and its average over the stroke disk, i.e. the local curvature of the surface.
It does so in a coordinate-invariant way.

Substituting into Eq.~\eqref{eq:coarse_grained_grinding_dynamics} and applying the same argument by symmetry for $z_T$:
\beq
\left\{
\bal
\partial_t z_B
&=
- \frac{1}{T_\text{grind}} 
\left[ z_B - z_T \right]
+
\frac{R_\text{stroke}^2}{8 T_\text{grind}}
\nabla^2 z_T\\
\partial_t z_T
&=
- \frac{1}{T_\text{grind}} 
\left[ z_T - z_B \right]
+
\frac{R_\text{stroke}^2}{8 T_\text{grind}}
\nabla^2 z_B\\
\eal
\right.
\eeq
Decoupling the variables with $\bar{z} \equiv (z_B + z_T)/2$ and $\delta z \equiv z_T - z_B$ leads to:
\beq
\left\{
\bal
\partial_t \bar{z}
&=
\underbrace{\frac{R_\text{stroke}^2}{8 T_\text{grind}}
\nabla^2 \bar{z}}
_{\text{diffusion}}\\
\partial_t \delta z
&=
\underbrace{- \frac{2}{T_\text{grind}} 
\delta z}
_{\text{exp. damping}}
\;
\underbrace{-
\frac{R_\text{stroke}^2}{8 T_\text{grind}}
\nabla^2 \delta z}
_{\text{``anti-diffusion''}}\\
\eal
\right.
\eeq

The mean surface $\bar{z}$ obeys a diffusion equation with diffusivity $\kappa \equiv R_\text{stroke}^2/(8 T_\text{grind})$: large-scale deviations from a sphere or plane diffuse away, faster for larger strokes and faster grinding.
A deviation of transverse size $L$ then relaxes in a diffusion time of order $L^2/\kappa$. The last to disappear is the largest one, spanning the whole mirror ($L \sim R_\text{mirror}$), which sets the spherizing time $T_\text{sphere} \sim (R_\text{mirror}/R_\text{stroke})^2\, T_\text{grind}$. Solving the diffusion equation for this longest-wavelength deviation (of wavelength $\lambda \simeq 4 R_\text{mirror}$, decaying as $(2/\pi^2)(\lambda/R_\text{stroke})^2\, T_\text{grind}$) fixes the numerical prefactor:
\beq
\boxed{
T_\text{sphere}
\equiv
\frac{32}{\pi^2} 
\left( \frac{R_\text{mirror}}{R_\text{stroke}} \right)^2 T_\text{grind}.
}
\label{eq:t_sphere}
\eeq
For the normal $1/3$-diameter stroke, $T_\text{sphere} \simeq 30\, T_\text{grind}$, consistent with experience: noticeable surface changes take $T_\text{grind} \lesssim$~hour, while spherization takes $T_\text{sphere} \sim$~tens of hours.

The mismatch $\delta z$ has two contributions. The exponential damping term drives the two surfaces to match on timescale $T_\text{match} \equiv T_\text{grind}/2$, a factor of two faster than $T_\text{grind}$ because both surfaces evolve simultaneously.
The anti-diffusion term reflects the fact that a point on $z_B$ facing a peak of $z_T$ feels the stroke-averaged $z_T$, which is less than the peak value; the correction is again captured by the Laplacian.
Taken at face value, this ``anti-diffusion'' term implies exponential growth of modes with $\lambda < (\pi/4)R_\text{stroke}$, but this is an artefact: the Laplacian approximation is only valid for $\lambda \gg R_\text{stroke}$, so the apparent instability lies outside the regime of validity of the equation.

\section{Destabilizing effect of gravity and edges}

\begin{figure}[htbp]
\centering
\includegraphics[width=0.9\columnwidth]{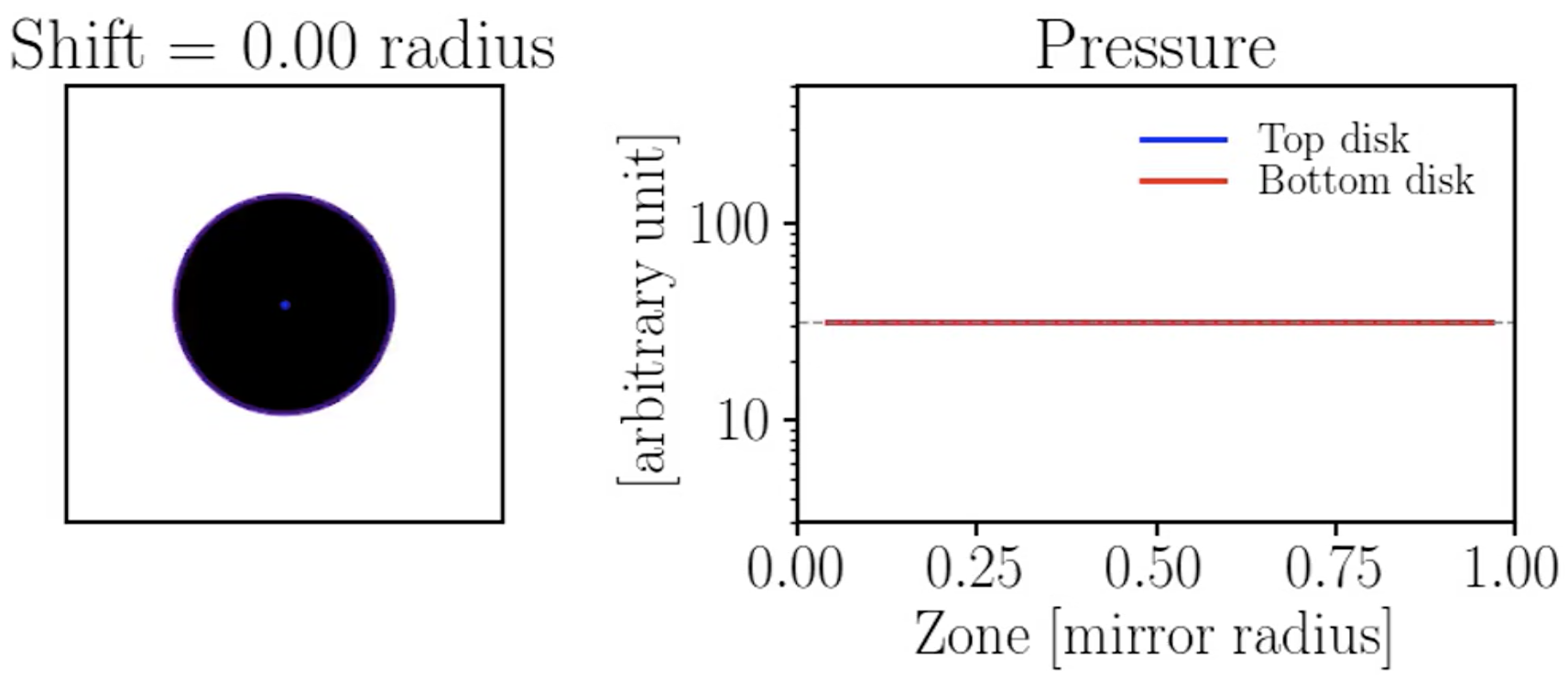}\\
\includegraphics[width=0.9\columnwidth]{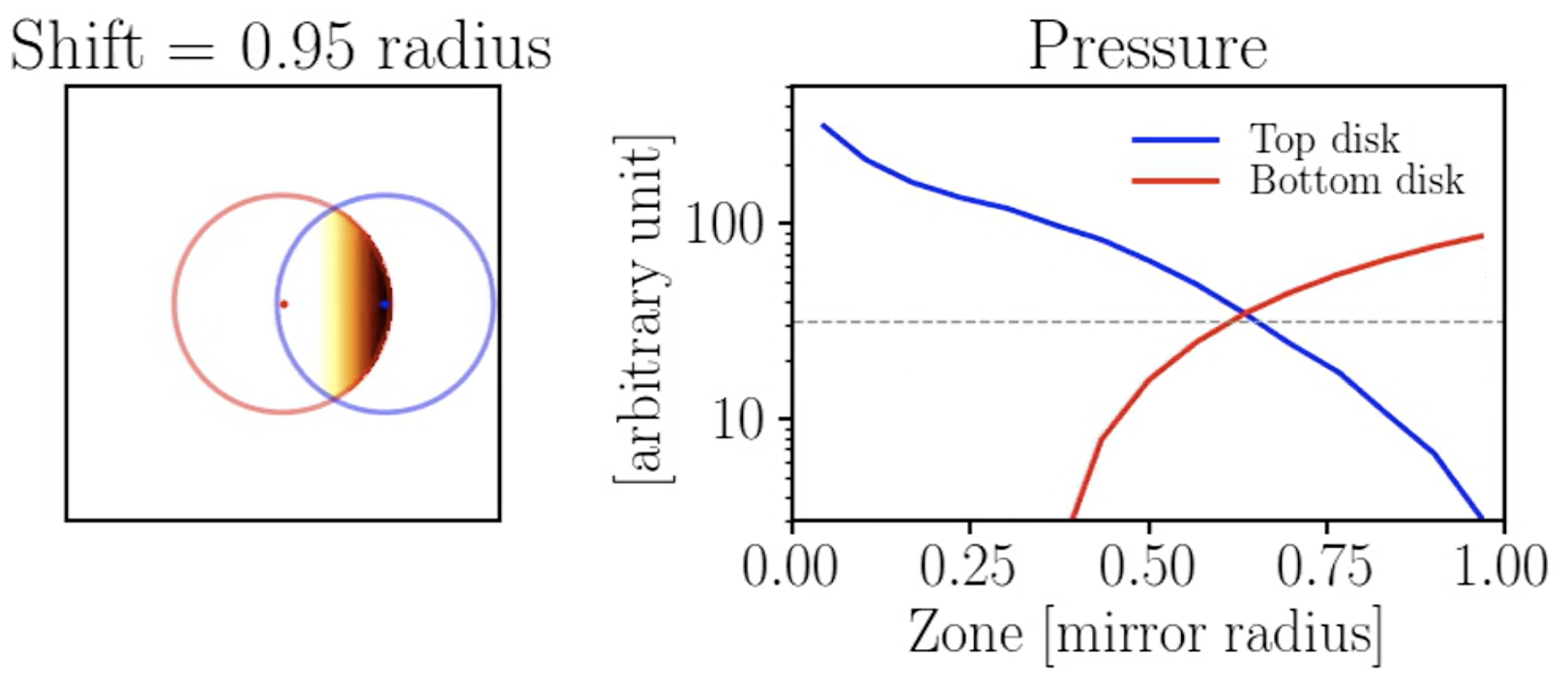}
\caption{
Without overhang (top), pressure is uniform and grinding is equal across both surfaces.
For extreme overhang (bottom), the pressure asymmetry is large: grinding efficiency at the edge and center can differ by large factors, with opposite trends for top and bottom disks.
Additional intermediate overhang amounts are shown in App.~\ref{app:edges_gravity}.
}
\label{fig:stroke_theory_result1}
\end{figure}

Without gravity, $z_T$ and $z_B$ are interchangeable and the equilibrium configurations (spheres and planes) are stable, as we have shown.
Gravity breaks this symmetry: as the stroke is performed, the top surface overhangs the bottom by a varying amount, generating gradients in the pressure field $P_\text{gravity}(\vec{x})$, illustrated in Figs.~\ref{fig:stroke_theory_result1}, \ref{fig:grinding_theory_fig}, \ref{fig:edges_gravity_full}.
We derive this pressure field in App.~\ref{app:edges_gravity}, assuming equilibrium surfaces and averaging over the stroke (details deferred there for clarity).
Two effects contribute:
\begin{itemize}
\item \textbf{Geometry of overlap (``edge effect''):} the fraction of each ring of radius $r$ where the two disks overlap, and thus where grinding occurs, drops toward the edge. This effect is symmetric for top and bottom surfaces.
\item \textbf{Pressure asymmetry from gravity:} overhang tilts the pressure field, breaking the top-bottom symmetry of the ring-averaged grinding action.
\end{itemize}
Fig.~\ref{fig:stroke_theory_result1} shows that large overhangs produce strong deviations from uniform grinding, driving the surfaces away from their equilibrium shapes and causing increasing mismatch between top and bottom.
Remarkably, for any non-zero overhang the top surface always sees excess pressure at its center and the bottom surface at its edges: gravity generically drives the top surface concave and the bottom convex, thus making planes unstable.
As Fig.~\ref{fig:stroke_theory_result1} shows, strokes with excursion comparable to the mirror radius produce extreme edge pressure on the bottom surface.

We define $T_\text{deform}$ as the timescale on which gravity and edge effects destabilize the equilibrium surface.
When $T_\text{sphere} \ll T_\text{deform}$, the surfaces remain matching spheres at all times, with gravity only adiabatically increasing their curvature: the spheres are stable, though their radius drifts slowly.
When $T_\text{deform} \ll T_\text{sphere}$, deformation outpaces spherization and the surfaces deviate from spheres and from each other.
From Fig.~\ref{fig:stroke_theory_result1}, this effect grows with overhang, so $T_\text{deform}$ decreases with $R_\text{stroke}$.
Since $T_\text{sphere} \propto R_\text{stroke}^{-2}$ follows the same trend but likely with a less steep scaling, we expect $T_\text{sphere} < T_\text{deform}$ for short strokes and the reverse for long ones.

\section{Confronting theory \& experience}

\subsection{Role of each stroke in mirror making}

Stroke length sets a trade-off: longer strokes spherize faster (larger $\kappa$) but also deform faster (smaller $T_\text{deform}$), so the ordering of $T_\text{sphere}$ and $T_\text{deform}$ fixes which stroke to use (Table~\ref{tab:strokes}, Fig.~\ref{fig:grinding_strokes}).
Rough grinding uses the long chordal stroke precisely because its short $T_\text{deform}$ digs the sagitta quickly; once the depth is reached, switching to the normal stroke reverses the ordering of the timescales, and the surfaces relax to matching spheres. The remaining strokes refine the surface: alternating MOT and TOT pins the radius of curvature  while smoothing the mirror, and the W-stroke departs from the sphere in a controlled way. Finally, flat mirror can be obtained by cycling three surfaces.
Thus we now understand the physics behind the recipe.
\begin{table*}[htbp]
\caption{Which stroke to use, set by the ordering of $T_\text{sphere}$ and $T_\text{deform}$.}
\label{tab:strokes}
\begin{tabular}{|p{0.13\textwidth} p{0.15\textwidth} p{0.17\textwidth} p{0.50\textwidth}|}
\hline
Stroke & Use & Timescales & Effect\\
\hline
Long chordal & Rough shaping:\newline dig sagitta. & $T_\text{deform}\ll T_\text{sphere}$ & Deformation outpaces spherizing, digging the mirror center and the tool edges into a rough concavity and convexity.\\
\hline
``Normal''\newline($1/3$~diameter) & Produce sphere. & $T_\text{sphere}\ll T_\text{deform}$ & The surfaces track their spherical equilibrium at all times; only the radius of curvature changes adiabatically.\\
\hline
Alternating MOT/TOT & Smooth sphere \newline at fixed radius \newline of curvature. & $T_\text{sphere}\ll T_\text{deform}$\newline Balances both signs & Gravity drives the top surface concave and the bottom convex; alternating mirror-on-top and tool-on-top cancels the two tendencies.\\
\hline
W-stroke\newline(mix chordal\newline$+$ normal) & Convert sphere \newline to paraboloid. & $T_\text{sphere}\sim T_\text{deform}$ & Blends a chordal component (a controlled departure from sphericity) with a normal one (which spreads it smoothly); it has no automatic attractor, so it requires iterating grinding and testing.\\
\hline
Three-surface cycling\newline(1/2, 2/3, 3/1) & Produce flat \newline secondary \newline mirrors. & $T_\text{sphere}\ll T_\text{deform}$\newline Stabilizes planes & Each surface spends equal time on top and bottom, and the cycle acts as a negative feedback loop that flattens all three disks.\\
\hline
\end{tabular}
\end{table*}

\subsection{Numerical estimates: grinding a 10'' mirror}
\label{sec:numerics}

Consider hand-grinding the 10'' or $25$~cm diameter mirror, whose grinding log is given in Table~\ref{tab:grit_log}.
The curvature desired ($f/5$) corresponded in this case to a 3~mm depth at the center of the mirror (or ``sagitta'').

The applied pressure is $P \sim 1$--$2$~kPa (both hands, $\sim 50$--$100$~N over $\sim 500$~cm$^2$), the relative speed is $v \sim 0.1$--$0.3$~m/s, and the grit size $d$ ranges from $\sim 200~\mu$m (coarsest grit) to a few microns (finest grit).
For loose-abrasive grinding of glass, Preston's coefficient grows roughly linearly with grit size in the brittle regime~\cite{lambropoulos1997}; scaling the micron-scale polishing value ($\sim 10^{-13}$~m$^2$/N) up to $200~\mu$m gives $\alpha \sim 10^{-10}$~m$^2$/N.
Preston's law then gives a removal rate $\alpha P v \sim$~a few $\mu$m/min at the coarse grit, so the $\sim 3$~mm central sagitta should be dug in $\sim 10$~hours, matching practice.

The e-folding time is $T_\text{grind}\sim d/(\alpha P v)$, the time to grind through about one grit-depth, of order tens of minutes to an hour.
Each grinding stage after the first removes the subsurface damage left by the previous, coarser grit, a depth $\sim d$; since the removal rate also grows with $d$, the two scalings cancel and $T_\text{grind}$ is nearly grit-independent.
This is why the successive fine-grit stages each last a comparable time rather than scaling with grit size.
The time spent at each grit stage in practice matches our estimate for a few e-folding times $T_\text{grind}$.

The first, sagitta-digging stage is different, since it aims to remove 3~mm, i.e. $\sim 15$ grit-depths for the 200~$\mu$m grit used.
Indeed, it is found to take roughly $\sim 15\,T_\text{grind} \sim 10$~hours.
The second stage, spherizing with the normal stroke, is instead controlled by the spherizing time of Eq.~\eqref{eq:t_sphere}: $T_\text{sphere} \simeq 30\, T_\text{grind} \sim 10$~hours for $T_\text{grind}$ of a few tens of minutes, again matching the observed duration and completing the explanation of the grinding log in Table~\ref{tab:grit_log}.

\begin{figure}[htbp]
\centering
\includegraphics[width=0.7\columnwidth]{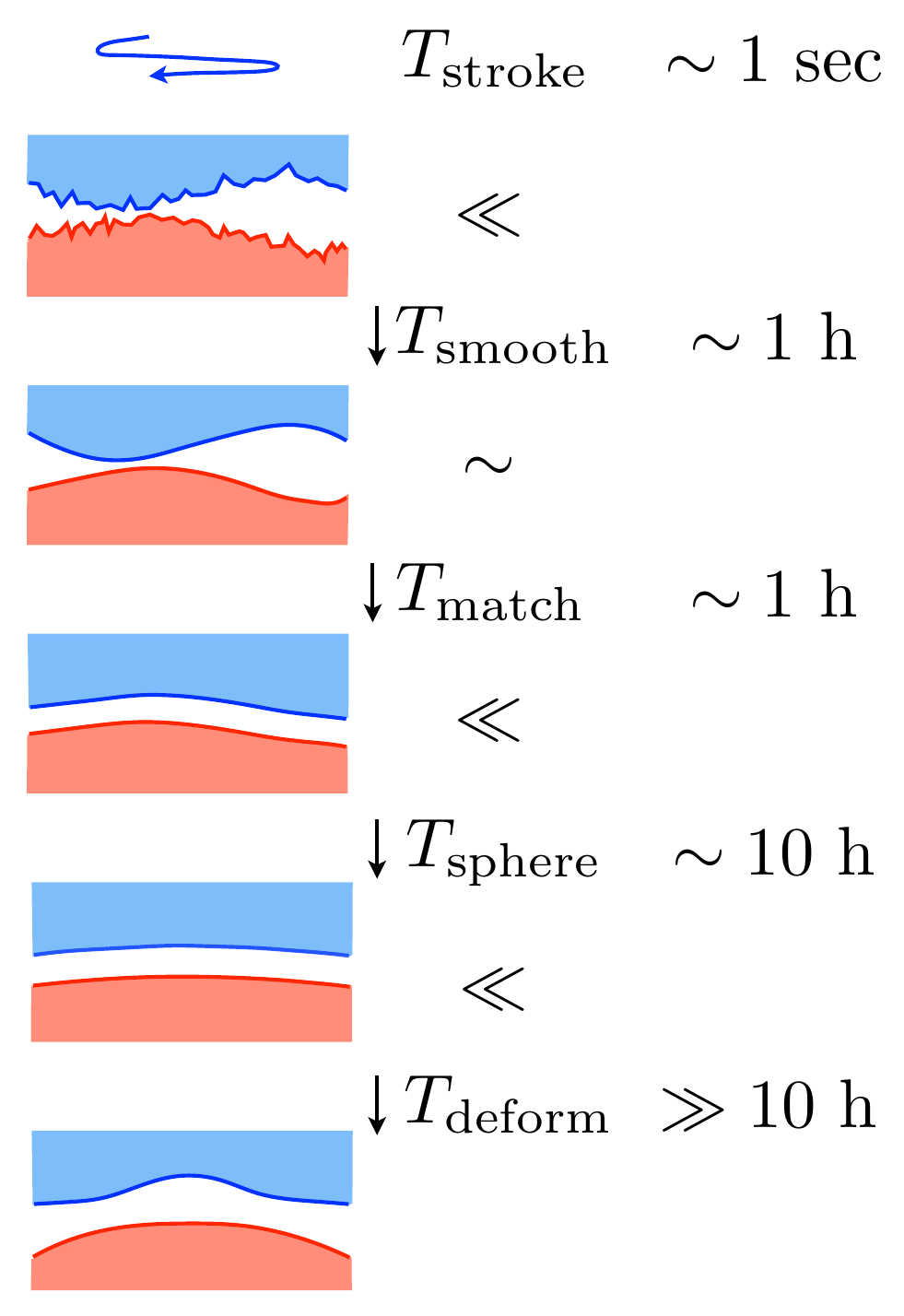}
\caption{
Key timescales and their hierarchy, leading to spheres as attractors of the grinding process.
Numerical values are estimates for a 10'' diameter mirror.
}
\label{fig:hierarchy_timescales}
\end{figure}

The characteristic timescales of the process, beyond $T_\text{grind}$, are summarized in Fig.~\ref{fig:hierarchy_timescales} for this 25~cm mirror.

\subsection{Comparison with rock tumbling: short strokes extend the spherizing time}\label{sec:rocks}

As a first test of this framework in a different regime, consider rock tumbling.
Rocks and grit rub against each other, with the container either rotating or vibrating.
The rock radius $R_\text{rock}$ plays the role of $R_\text{mirror}$, and the collision or sliding excursion the role of $R_\text{stroke}$.
Practitioners observe that (i) both rotary and vibratory tumbling produce shiny rocks; (ii) vibratory tumbling does so faster; (iii) rotary tumbling produces more rounded rocks; and (iv) neither produces spherical rocks on reasonable timescales.
Our framework explains all four.

In rock tumbling, the stroke excursion is very short ($\sim 1$~mm) compared to the rock size ($\sim 1$~cm): the rocks mostly collide rather than smoothly slide.
Irregularities on scales below the stroke ($\lesssim$~mm) are therefore damped on timescale $T_\text{smooth} = T_\text{grind}$, producing the observed shininess in a few weeks (observations i--ii).
Vibratory tumbling has a higher collision frequency, hence shorter $T_\text{grind}$, and thus smooths faster than rotary tumbling.
Rotary tumbling, by contrast, involves longer sliding contacts and thus a larger effective stroke excursion, damping irregularities on larger scales and producing more rounded edges (observation iii).

In both cases, the large ratio $(R_\text{rock}/R_\text{stroke})^2 \sim 100$ pushes $T_\text{sphere}$ from Eq.~\eqref{eq:t_sphere} to $\sim$~years, far beyond the typical few weeks of tumbling, so spherical rocks are not expected (observation iv).

These findings connect with the literature on rock abrasion under isotropic bombardment \cite{firey1974, bloore1977, domokos2012, havasi2023, priour2020}.
Ref.~\cite{bloore1977} shows that the linearized abrasion equation expanded around a sphere reduces to a diffusion equation for surface perturbations.
The present paper arrives at the same diffusion equation by a different route, from Preston's law and the stroke geometry rather than isotropic bombardment.
It also extends it to the two-body mirror-tool problem, which introduces matching dynamics with no analog in single-body pebble models.

Sanding a small flat object is the opposite regime. Here the stroke excursion typically exceeds the object size, so $T_\text{deform}$ becomes comparable to $T_\text{grind}$: deformation is essentially immediate, and the edges inevitably round. This is why it is so hard to keep a small surface flat by hand, and why the plane, unlike the sphere, is not an attractor of the process.

\section{Conclusion}

Starting from Preston's law of abrasion, we have derived a first-principles explanation for why rubbing two glass disks together spontaneously produces matching spherical surfaces.
The key insight is a hierarchy of timescales:
\beq
\boxed{
T_\text{stroke} \ll T_\text{match} \sim T_\text{smooth} = T_\text{grind} \ll T_\text{sphere} \ll T_\text{deform},
}
\eeq
illustrated in Fig.~\ref{fig:hierarchy_timescales}.
The first inequality ($T_\text{stroke} \ll T_\text{grind}$) places us in the grinding regime, where each stroke has only a small cumulative effect, allowing a stroke-averaged secular analysis.
The middle inequalities reflect that smoothing and matching, both exponential in time, occur on the fast timescale $T_\text{grind}$ (or $T_\text{grind}/2$), while spherization is slower by a factor $(R_\text{mirror}/R_\text{stroke})^2$ because it is a diffusion process.
The final inequality ($T_\text{sphere} \ll T_\text{deform}$) ensures that surfaces reach their spherical equilibrium faster than gravity and edge effects can destabilize them; it is satisfied by choosing sufficiently short strokes.
In this regime, the sphere's radius of curvature simply varies adiabatically: the surfaces are always nearly spherical, just becoming progressively more curved.

It is remarkable that a stroke exists that satisfies these conditions.
Practice shows that the normal stroke ($1/3$-diameter travel) sits in this sweet spot.
Longer strokes would speed up smoothing but shrink $T_\text{deform}$, eventually violating the final inequality and producing aspherical, mismatched surfaces, as in rough grinding with the chordal stroke, where this is deliberate.
There is enough wiggle room in the stroke length that achieving smooth, matching spheres does not require precise control and can be done manually, which explains the robustness of the recipe in practice.
Conversely, the hierarchy also implies that spherization is inevitably slow: it takes many strokes to match and smooth the surfaces, and many more to diffuse large-scale deviations away.

The same framework unifies two other everyday phenomena.
In rock tumbling, the very short stroke excursion ($R_\text{stroke} \sim 1$~mm against rock size $\sim 1$~cm) produces a spherizing time orders of magnitude longer than the smoothing time, explaining why tumbled rocks are shiny but not spherical.
This also explains why rotary tumbling, with its longer effective stroke, rounds edges more than vibratory tumbling.
In sanding small flat objects on the other hand, the stroke excursion typically exceeds the object size, making $T_\text{deform}$ comparable to $T_\text{grind}$: deformation is essentially immediate, which is why flat surfaces inevitably round.

Our model neglects several effects: friction and horizontal forces from the stroking motion (which alter the pressure field, especially for radial strokes); the operator's hand pressure distribution; and the viscoelastic pitch tool used in polishing (post grinding), which introduces an additional timescale $T_\text{viscoelastic}$ that we expect must satisfy $T_\text{sphere} \ll T_\text{viscoelastic} \ll T_\text{deform}$ for the polishing phase to work correctly.
That said, the main results, planes and spheres as attractors, exponential matching, diffusive spherization, follow from symmetry and timescale separation rather than modeling details, suggesting they are robust to these idealizations. A more systematic empirical test of the predicted timescales, for example with instrumented hand grinding, would be a valuable complement to the order-of-magnitude checks presented here.

Finally, the testing of the finished mirror surface, confirming smoothness and figure accuracy using nothing more than a light source and a razor blade, is as physically rich as the grinding itself, and equally accessible in the classroom.
We leave this to a future article.

\begin{acknowledgments}

The author is deeply grateful to the members of the Telescope Makers' Workshop at Chabot Space \& Science Center in Oakland, CA, who generously shared their time and expertise in teaching him how to grind, polish, and figure a telescope mirror (a 25 cm (10 inch) diameter, f/5 paraboloidal mirror). 
Special thanks go to Gert Gottschalk, Richard Ozer, Dave Barosso, David Lawson, Phil Okada, Henner Zeller, Rolph, Nate, and Alan. 
The author also thanks Patrick Jelinsky for lending him a wire tester, which helped in understanding the effect of various strokes on the mirror figure,
and Megan deBettencourt for her support throughout the mirror-making process. Finally, the author is grateful to the editor and reviewers for their thoughtful feedback and suggestions, which improved the clarity of the paper.
The author performed this work independently; it does not represent the views of SLAC National Accelerator Laboratory or the U.S. Department of Energy.

\end{acknowledgments}

\bibliography{references}

\newpage   
\appendix

\section{Pressure profile from edges and gravity}
\label{app:edges_gravity}

We compute the radial grinding efficiency profile for two circular disks of radius $R$, offset by $s \in [0, R]$, assuming a translational stroke at uniform speed.
The key quantity is the pressure field in the overlap region, averaged circularly over each disk.

To close the otherwise hyperstatic system, we treat the top disk as rigid and the bottom disk as elastic, so that the bottom disk's deformation is an affine function of position, determined by force and moment balance.
The true pressure distribution depends on the relative hardness of the two materials (and, in polishing, on the viscoelastic pitch), making a fully realistic model highly material-specific.
Our assumption captures the generic feature that matters: a pressure gradient develops to balance the gravitational torque, driving the differential grinding between center and edge.
We adopt the full-contact approximation throughout, which becomes inaccurate at extreme overhangs where the tilted top disk lifts off part of the overlap area.

\paragraph*{Geometry and Contact Area}

Let $R$ be the disk radius and $s \in [0,R]$ the center offset.
\begin{figure}[h]
\centering
\includegraphics[width=0.8\columnwidth]{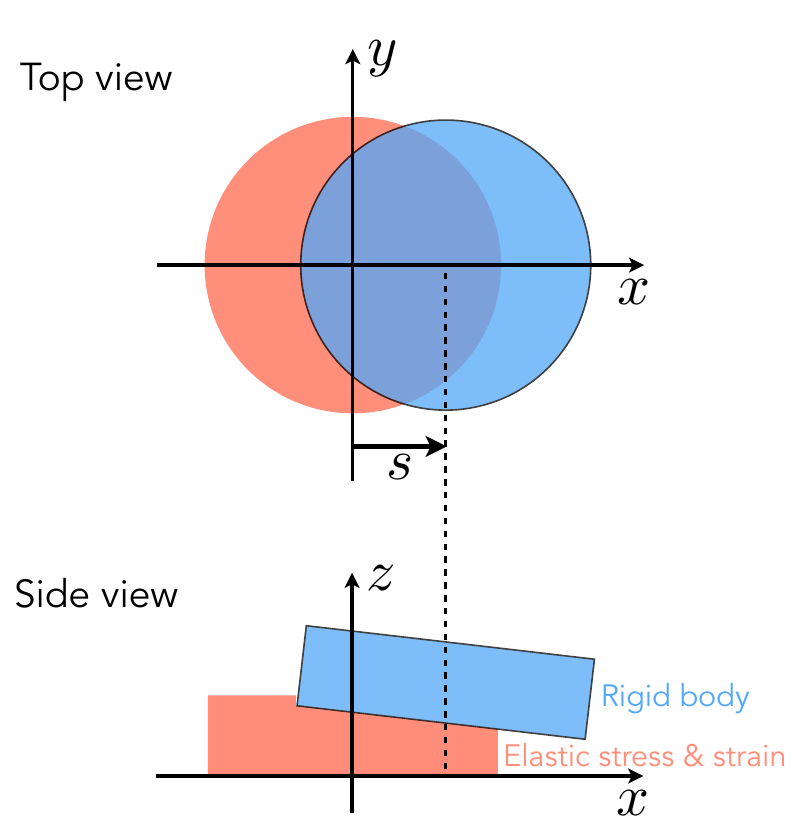}
\caption{
Geometry of the overhang.
The pressure field in the overlap region is computed by treating the bottom surface as elastic and the top surface as perfectly rigid, avoiding the hyperstatic problem while retaining the generic pressure gradient needed to balance the gravitational torque.
}
\label{fig:grinding_theory_fig}
\end{figure}

The upper boundary of the overlap region is:
\beq
y(x) =
\begin{cases}
\sqrt{R^2 - x^2}, & \frac{s}{2} \le x \le R, \\
\sqrt{R^2 - (x - s)^2}, & s - R \le x \le \frac{s}{2},
\end{cases}
\eeq
with area element $dA(x) = 2y(x)\,dx$.

\paragraph*{Force Model}

We assume a linear pressure profile{\footnote{This tilted linear pressure distribution is the same one used to model edge effects in computer-controlled optical surfacing~\cite{kim2009, jones1977, aspden1972}, which independently supports the pressure model adopted here.}}, enforcing positivity and confinement to the overlap region via indicator functions:
\beq
f(x,y) = (ax + b)\,\mathbf{1}_{\{ax + b \ge 0\}} \, \mathbf{1}_{\{s - R \le x \le R\}}.
\eeq
Under the full-contact approximation this simplifies to:
\beq
f(x,y) = (ax + b)\,\mathbf{1}_{\{s - R \le x \le R\}}.
\eeq

\paragraph*{Balance of Forces}

Requiring the integrated pressure to equal the total weight $F$:
\begin{equation}
a 
\underbrace{
\left[ \int_{s-R}^{R} dx\, 2y(x)\, x \right] 
}_A
+ b 
\underbrace{
\left[ \int_{s-R}^{R} dx\, 2y(x) \right] 
}_B
= F.
\tag{force balance}
\end{equation}

\paragraph*{Balance of Moments}

Requiring zero net moment about $x = s$:
\begin{equation}
a 
\underbrace{
\left[ \int_{s-R}^{R} dx\, 2y(x)\,x(x - s) \right] 
}_C
+ 
b 
\underbrace{
\left[ \int_{s-R}^{R} dx\, 2y(x)(x - s) \right] 
}_D
= 0.
\tag{moment balance}
\end{equation}

This yields the matrix system:
\begin{equation}
\begin{pmatrix} A & B \\ C & D \end{pmatrix}
\begin{pmatrix} a \\ b \end{pmatrix}
=
\begin{pmatrix} F \\ 0 \end{pmatrix},
\end{equation}
with solution:
\begin{equation}
\boxed{
\begin{pmatrix} a \\ b \end{pmatrix}
= \frac{1}{AD - BC}
\begin{pmatrix} D & -B \\ -C & A \end{pmatrix}
\begin{pmatrix} F \\ 0 \end{pmatrix}.
}
\end{equation}

\begin{figure*}[htbp]
\centering
\includegraphics[width=0.4\textwidth]{figures/grinding_theory/result1.png}\\
\includegraphics[width=0.4\textwidth]{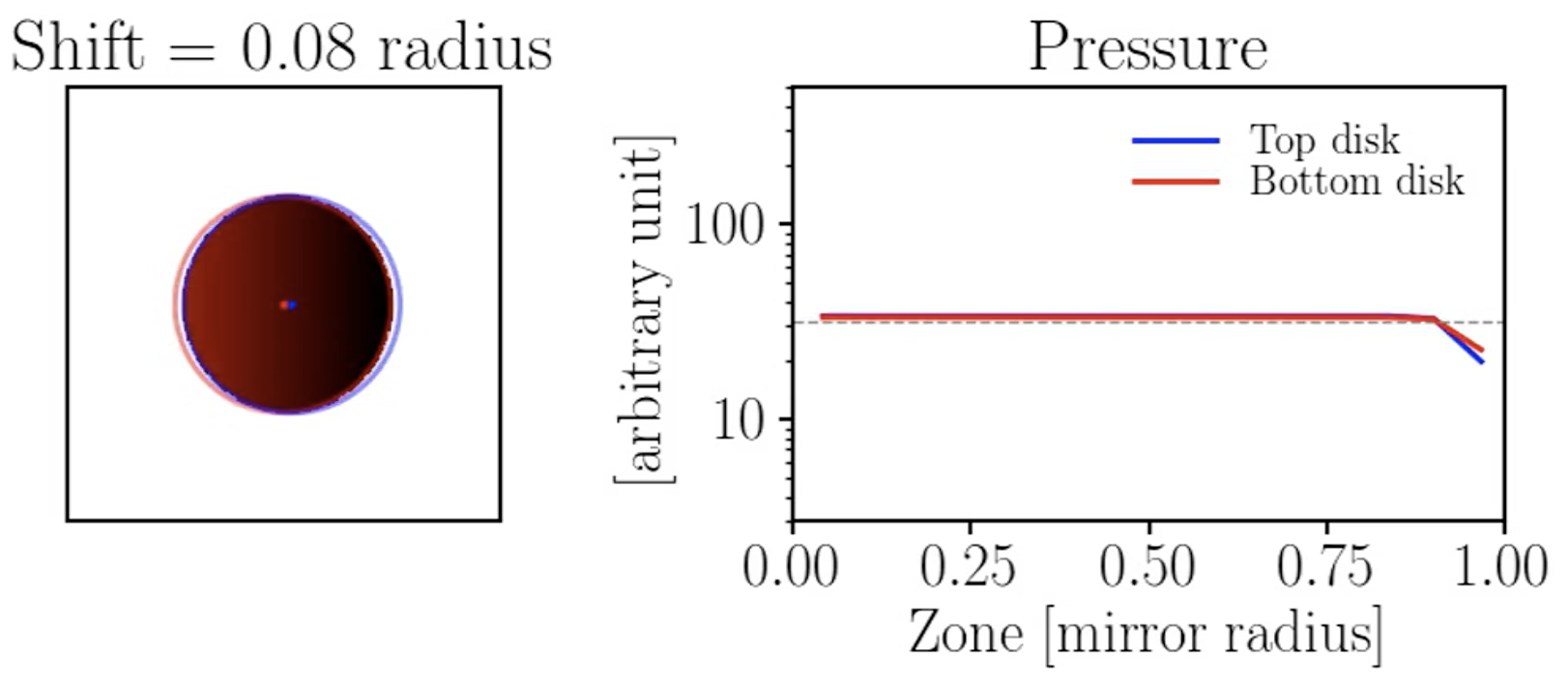}\\
\includegraphics[width=0.4\textwidth]{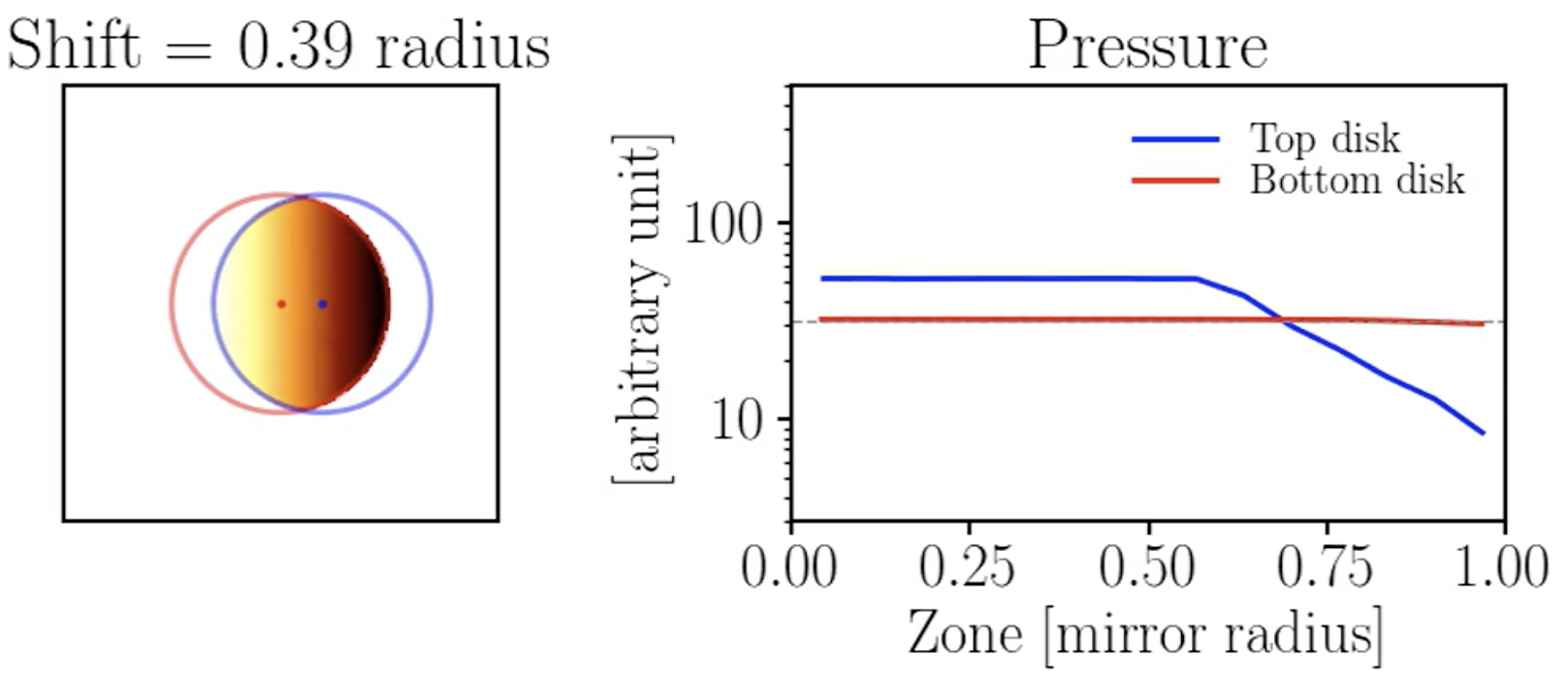}\\
\includegraphics[width=0.4\textwidth]{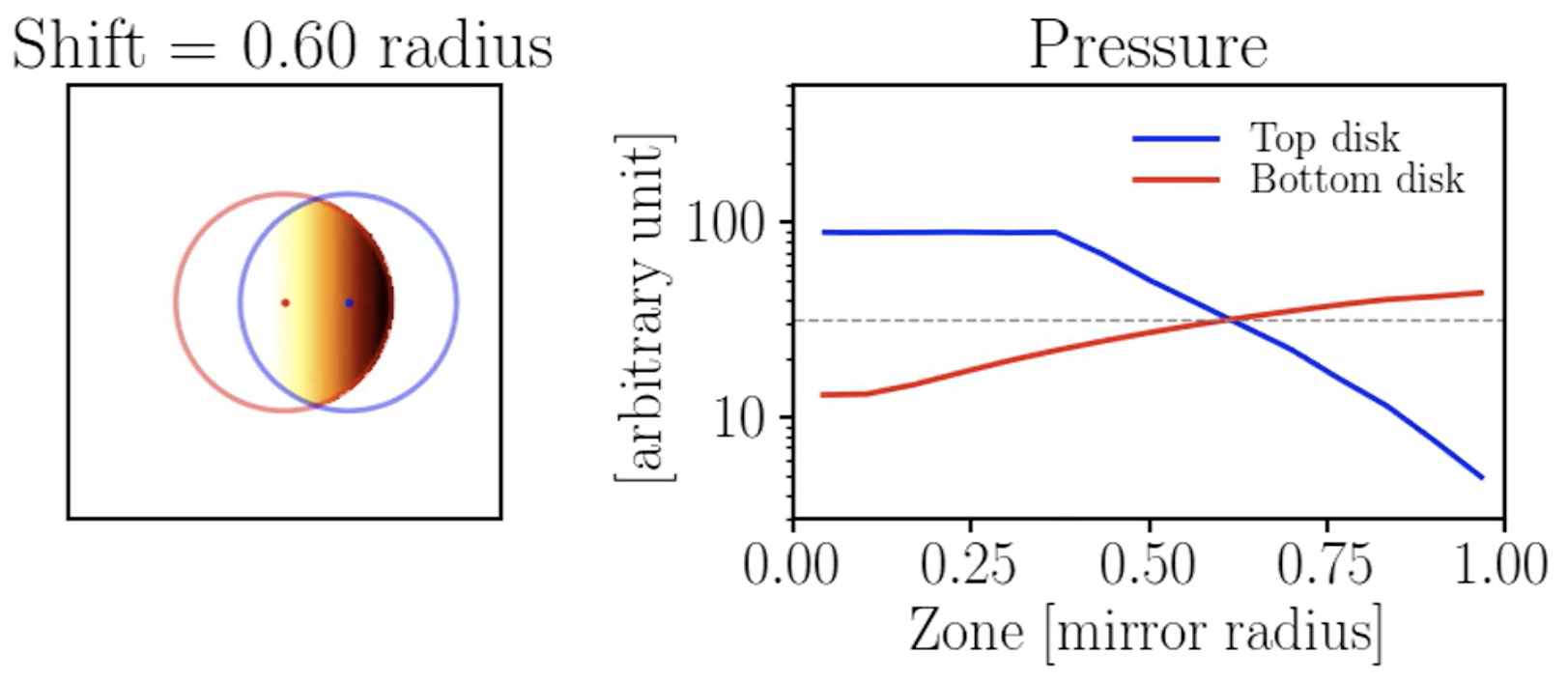}\\
\includegraphics[width=0.4\textwidth]{figures/grinding_theory/result5.png}
\caption{
Ring-averaged grinding efficiency as a function of radial position, for both disks, at five overhang values.
\textit{No overhang (row 1):} pressure is uniform; grinding efficiency is equal and flat across both disks, consistent with weight divided by area.
\textit{Small overhang (row 2):} the pressure asymmetry is negligible; only the geometric overlap matters, reducing efficiency symmetrically toward the edge of both disks.
\textit{Sizable overhang (row 3):} a significant pressure gradient develops. For the top disk, the gradient averages out across each fully-overlapping ring, leaving its efficiency flat; for the bottom disk, the gradient partially cancels the geometric drop, making its efficiency more uniform across the whole disk.
\textit{Large overhang (row 4):} the pressure asymmetry dominates. The top disk retains uniform efficiency only for rings fully within the (now narrower) overlap zone. For the bottom disk, edge rings --- even partially overlapping --- are ground harder than the center.
\textit{Extreme overhang (row 5):} the pressure gradient is extreme and the full-contact approximation breaks down. The grinding efficiencies of top and bottom disks are opposite and strongly non-uniform, with center-to-edge ratios reaching large factors.
}
\label{fig:edges_gravity_full}
\end{figure*}

The pressure field $f(x,y) = (ax + b)\,\mathbf{1}_{\{s - R \le x \le R\}}$ is then fully determined.
For uniform stroke speed, the grinding action on each radial ring is the ring-averaged pressure, shown in Fig.~\ref{fig:edges_gravity_full} for five representative overhangs.

\end{document}